\documentclass[twocolumn,showpacs,preprintnumbers,amsmath,amssymb]{revtex4} 
\usepackage{hyperref}
\usepackage{amsfonts}
\usepackage{epsf}
\usepackage{t1enc}
\usepackage[latin2]{inputenc}

\usepackage{graphicx}
\usepackage{dcolumn}
\usepackage{bm}

\DeclareFontFamily{OT1}{rsfs}{} \DeclareFontShape{OT1}{rsfs}{m}{n}{
<-7> rsfs5 <7-10> rsfs7 <10-> rsfs10}{}
\DeclareMathAlphabet{\mycal}{OT1}{rsfs}{m}{n}

\newtheorem{condition}{Condition}[section]
 
\begin{document}

\title{ Massive fields tend to form highly
    oscillating \\ self-similarly expanding shells}

\author{Gyula Fodor}
\email{gfodor@rmki.kfki.hu}
\author{Istv\'{a}n R\'{a}cz}%
\email{istvan@sunserv.kfki.hu}
\affiliation{%
MTA KFKI, Részecske- és Magfizikai Kutatóintézet\\
H-1121 Budapest, Konkoly Thege Miklós út 29-33.\\
Hungary\\}%

\date{\today}

\begin{abstract}{\footnotesize
The time evolution of self-interacting spherically symmetric scalar
fields in Minkowski spacetime is investigated based on the use of
Green's theorem. It is shown that a massive Klein-Gordon field can be
characterized by the formation of certain expanding shell structures
where all the shells are built up by very high frequency oscillations.
This oscillation is found to be modulated by the product of a simple
time decaying factor of the form $t^{-{3}/{2}}$ and of an essentially
self-similar expansion. Apart from this self-similar expansion the
developed shell structure is preserved by the evolution. In
particular, the energy transported by each shell
appears to be time independent.}
\end{abstract}

\maketitle

\parskip 5pt

\small

\section{Introduction}
Because of its importance in
black hole physics, the late time evolution of various fields on
a fixed spherically symmetric static asymptotically flat background
spacetime is of obvious interest. The physical mechanism by
which a scalar field in such an  asymptotically flat spacetime is
radiated away has been extensively investigated by means of analytical
and numerical  techniques. First, the case of  massless fields was
considered \cite{price,leaver,gpp,gpp2,buor} later the evolution of
self-interacting (massive) fields was also investigated
\cite{chop,hodpiran,koto,pl}. In both cases it appeared that
the late time evolution of scalar fields is dominated by an inverse
power-law  behavior. In particular, it has been found (see,
e.g., \cite{hodpiran}) that at a fixed radius, in the intermediate
asymptotic region, each multipole moment $\Phi_l$ of a
self-interacting scalar field $\Phi$ with a mass parameter $m$ evolves
according to the oscillatory inverse power-law behavior
\begin{equation}
\Phi_l \sim t^{-l-\frac32}\cos\left[mt-\left(\frac 12 l-\frac
\pi4\right)\right] \label{af1}
\end{equation}
in the limit $t\gg \frac1m$.

It is a common feature of all of the previous investigations that the
late time behavior of the scalar field is monitored only through the
investigation of the field variable at a fixed radius. This might
explain why the self-similar part of the decay rate, described in
detail below,
was not noticed in either of these considerations.

There has been only very limited attention paid to the study of the
behavior of massive fields at null infinity. Among the very few
relevant investigations the most important one 
concerning this problem is due to Winicour. He showed  
that the time evolution of massive fields in case of initial data of
compact support necessarily yields ${\mathcal{O}}
(1/r^\infty)$ asymptotic behavior at null infinity \cite{wini}. 

Having all the above results, it might be somewhat surprising that this
paper is dedicated to the investigation of the dynamical properties of
a massive spherically symmetric Klein-Gordon (KG) field on the 
Minkowski spacetime. 
In fact the necessity of studying this problem arose
during the examination of the evolution of a more complicated dynamical
system.  We started by investigating the evolution of ``excited
magnetic monopoles''. More precisely, in \cite{fr} the time evolution
of spherically symmetric Yang-Mills--Higgs systems is  numerically
 considered on a
fixed Minkowski background spacetime. To have a
computational grid covering the full physical spacetime,  the
techniques of conformal compactification, along with the hyperboloidal
initial value problem, were used. In this way it was possible to
study the asymptotic behavior of the fields close to future null
infinity for considerably long physical time intervals. The numerical
simulations indicated that when massive fields are included
certain expanding shell structures form, where all the
shells are built up by very high frequency oscillations. Since the
maximal frequency of these oscillations is growing in (both the
physical and nonphysical)
time, the associated wavelengths reach the size of any fixed
equal distance grid spacing specified in advance in the
calculations. This may be expected to yield a significant
contamination of the numerical data. Therefore, after only the
first dozens of runs, we suspected that the formation of the above
mentioned high frequency oscillations and expanding shell structures
was due to numerical noise. Nevertheless, convergence tests,
along with  the monitoring of the energy conservation, made it apparent
that we have to take this as a physical phenomenon.  The next
natural question was whether the existence of these structures is due
to the nonlinearities associated with a coupled  Yang-Mills--Higgs
system or whether the same effect may also occur in the case of a 
simple linear
field as well. This issue is of obvious physical importance, because
far from the center of symmetry the field values are expected to
differ only slightly from that of the static ``background'' monopole
solution; moreover, the linearized equations, relevant for the
asymptotic region, take the form of a pair of uncoupled massive and
massless KG fields. Our aim in this paper, in addition to presenting 
the main
features of  this phenomenon, is to demonstrate that they are not the
nonlinear effects that are responsible for the appearance of
expanding highly oscillating shell structures, since the same features
already appear in the evolution of massive scalar fields.

To simplify the physical setting, in this paper only the evolution of a
spherical symmetric massive KG field  is investigated. It is shown
that for essentially arbitrary initial data with compact support the
evolution can be characterized by the formation of expanding shells
built up by very high  frequency oscillations. 
As the time passes the maximal physical frequency of the oscillations
forming the outer shells increases without any upper bound and thereby
more and more shells become visible. There is also an
argument presented explaining the qualitative and quantitative
properties of the underlying physical process. We found that the exact
time evolution of an initial data function with compact support can 
always be
approximated by a simple expression to a very high precision in a
considerably large portion of the Cauchy development. It turns out
that the evolution of such a massive KG field yields high frequency
oscillations modulated by the product of a simple time decaying factor
of the form $t^{-{3}/{2}}$ and of an essentially self-similar
expansion. 
Despite the massive character of the KG field, the ``edge'' of this
self-similar expansion moves with the velocity of light. 
This behavior of the KG field has also been justified for
the full (nonapproximate) description by numerical integration based
on Green's theorem. As far as the authors know, the appearance of this
scaled down  self-similar feature of the modulation of the high
frequency mode has not been noticed yet, and it seems to be of physical
interest in its own right. The developed shell structures are found to
be stable, i.e., they seem to have their own lasting individual
existence. In particular, the total energy which can be associated
with each of them and which is transported by them is apparently
conserved during the evolution despite the underlying expansion.

The plan of this paper is as follows. In Sec.\ \ref{pre} we briefly
describe the investigated physical system. In Sec.\ \ref{vac} the time
evolution of the deformation of a vacuum configuration is
considered. The exact and approximated field values are determined by
making use of Green's theorem and a suitable approximation
process. The corresponding  analysis is outlined in Sec.\ \ref{nvac} for
temporarily static initial data. In Sec.\ \ref{en} further
characterization of the observed expanding shell structures is
presented in terms of the energy density and energy current density
profiles.  We conclude in Sec.\ \ref{fine} with a brief summary of our
results along with some of their implications.

\section{Preliminaries}\label{pre}
\setcounter{equation}{0}

This section is to recall some of the basic notions and results we
shall use in considering the evolution of a massive KG field $\Phi$
satisfying 
\begin{equation}
\Box\Phi+m^{2}\Phi=0 \label{kggen}
\end{equation}
on a fixed Minkowski background.
Given a point source at a point with Minkowski coordinates 
${x'}^{\alpha},$ the
retarded Green's function yields the generated field value at the point
$x^{\alpha}$ as
\begin{equation}
G(x;x')=
\frac{1}{2\pi}H(x^{0}-{x'}^{0})\left[  \delta
(\lambda)-\frac{m}{2\sqrt{\lambda}}H(\lambda)J_{1}(m\sqrt{\lambda})\right], 
\label{gf1}
\end{equation}
where $H$ is the Heaviside function, $\delta$ is the Dirac delta function,
$J_{1}$ is the Bessel function of the first kind of order one, and
$\lambda$ is the square of the Lorentz
distance separating the two points, i.e.
\begin{equation}
\lambda=(x^{\alpha}-{x'}^{\alpha})(x_{\alpha}-x'_{\alpha}). \label{lamg}
\end{equation}
Using the Bessel function relationships
$J_{1}(\xi)=-\frac{\partial}{\partial\xi}J_{0}(\xi)$ and $J_{0}(0)=1$, the
Green's function can also be written in the alternative form
\begin{equation}
G(x;x')=\frac{1}{2\pi}H(x^{0}-{x'}^{0})\frac{\partial}
{\partial\lambda}\left[  H(\lambda)J_{0}(m\sqrt{\lambda})\right].
\label{gf2}
\end{equation}
Given the field $\Phi$ and its normal derivative
$n^\epsilon\partial_\epsilon\Phi$ on a spacelike hypersurface
$\Sigma$, by virtue of Green's theorem, in the causal future of 
$\Sigma$ the field value can be expressed as
\begin{equation}
\Phi(x)= 
{\displaystyle\int\limits_{\Sigma}}
d\Sigma\big[  G(x;x')n^\epsilon\frac{\partial}{\partial {x'}^\epsilon}
\Phi(x')-\Phi(x')n^\epsilon\frac{\partial}{\partial {x'}^\epsilon}
G(x;x')\big], \label{grth}
\end{equation}
where $n^\epsilon$ denotes the future pointed unit normal vector field
on $\Sigma$. 
Whenever $\Sigma$ is chosen to be a $t'=const.$ flat hypersurface of
Minkowski spacetime the operator $n^\epsilon\frac{\partial}{\partial
  {x'}^\epsilon}$ is simply 
the partial derivative $\partial_{t'}$.

The first term in Eq.\ (\ref{grth}) gives the contribution yielded by the
excitation $(\partial_t\Phi)_\Sigma$ of a vacuum initial data function
with
$\Phi|_\Sigma=0$, while the second term can represent the evolution of a
temporarily static configuration, i.e., whenever 
$(\partial_t\Phi)_\Sigma=0$ but $\Phi|_\Sigma$ is nonvanishing.
As we will see in the following sections it is easier to consider
the evolution of these two special types of initial data settings
separately. Note that because of the linearity of the system the
evolution of a general initial data specification is simply yielded by the
superposition of these two types.

Whenever both the background and the KG field are
spherically symmetric it is advantageous to use spherical coordinates
$(t,r,\vartheta,\varphi)$. Then the line element reads
\begin{equation}
ds^{2}=dt^{2}-dr^{2}-r^{2}\left(  d\vartheta^{2}+\sin^{2}\theta\,d\varphi
^{2}\right),  \label{ds}
\end{equation}
while Eq.\ (\ref{kggen}) takes the form
\begin{equation}
\frac{\partial^{2}\Phi}{\partial t^{2}}-\frac{\partial^{2}\Phi}{\partial
r^{2}}-\frac{2}{r}\frac{\partial\Phi}{\partial r}+m^{2}\Phi=0
\end{equation}
for the field variable $\Phi=\Phi(t,r)$.
The associated energy density of the field is
\begin{equation}
\varepsilon=\frac{1}{2}\left[  \left(  \frac{\partial\Phi}{\partial t}\right)
^{2}+\left(  \frac{\partial\Phi}{\partial r}\right)  ^{2}+m^{2}\Phi
^{2}\right],\label{ed}
\end{equation}
while the outgoing energy current density is
\begin{equation}
S=-\left(  \frac{\partial\Phi}{\partial t}\right)  \left(  \frac{\partial\Phi
}{\partial r}\right).\label{ec}
\end{equation}

In analyzing the behavior of the spherical field $\Phi$, one can
assume, without loss of generality, that the observer lies on the axis
of rotation associated with the spherical coordinate system, 
with coordinates $x^{\alpha}=(t,r,0,0)$. 
Denoting the coordinates of the source point as
${x'}^{\alpha}=(t',r',\vartheta',\varphi')$, the relation (\ref{lamg})
takes the form
\begin{equation}
\lambda=(t-t')^{2}-{r'}^{2}-r^{2}+2r' r\cos\vartheta'\ . \label{lam}
\end{equation}

\section{Deformation of a vacuum configuration}\label{vac}
\setcounter{equation}{0}

We start off by choosing the $t=0$ hypersurface as our initial
data surface $\Sigma_0$ and moreover
assume that $\Phi|_{\Sigma_0}=0$ but $(\partial_t\Phi)|_{\Sigma_0}=
\dot{\Phi}_\circ(r)$ where $\dot{\Phi}_\circ:[0,\infty)\rightarrow
  \mathbb{R}$ is a sufficiently regular function of $r$. To get the
  field values we  
have to evaluate only the first term in Eq.\ (\ref{grth}). By making use of
Eq.\ (\ref{gf2}) at a point of the axis of rotation with coordinates
$x^{\alpha}=(t,r,0,0)$ with $t>0$, we get
\begin{eqnarray}
\Phi(t,r)\negthinspace &=&\negthinspace \int\limits_{0}^{\infty}\negthinspace
dr'\negthinspace\int\limits_{0}^{\pi}\negthinspace 
d\vartheta'\negthinspace\int\limits_{0}^{2\pi}\negthinspace d\varphi'
\dot{\Phi}_\circ(r')G(t,r;r',\varphi',\vartheta')
    {r'}^{2}\negthinspace\sin\vartheta' 
\nonumber\\
\negthinspace &=&\negthinspace \int\limits_{0}^{\infty}\negthinspace dr'
{r'}^{2}\dot{\Phi}_\circ(r')\negthinspace
\int\limits_{0}^{\pi}\negthinspace d\vartheta'  
\frac{\partial}{\partial\lambda}\left[  H(\lambda)J_{0}(m\sqrt{\lambda
})\right]  \sin\vartheta'.\nonumber\\
\end{eqnarray}
By virtue of Eq.\ (\ref{lam}), $\frac{\partial}{\partial \lambda}=
\left(\frac{\partial \lambda}{\partial {\vartheta'}}\right)^{-1}
\frac{\partial }{\partial {\vartheta'}}= -\frac{1}{2 r' r \sin \vartheta'}
\frac{\partial }{\partial {\vartheta'}}$ holds. 
This makes it possible to evaluate the $\vartheta'$ integral, which
provides 
\begin{eqnarray}
& &\hspace{-1.5cm}
\Phi(t,r) =
\frac{1}{2r}\int\limits_{0}^{\infty}dr'\ r' \dot{\Phi}_\circ(r')\left[
H(\lambda_{0})J_{0}(m\sqrt{\lambda_{0}}) \right. \nonumber\\ &
&\hspace{2.5cm} \left. -H(\lambda_{\pi})J_{0} 
(m\sqrt{\lambda_{\pi}})\right], 
\label{Phiv}
\end{eqnarray}
where $\lambda_{0}$ and $\lambda_{\pi}$ are the values of $\lambda$ at
$\vartheta'=0$ and $\vartheta'=\pi,$ i.e.,
\begin{equation}
\lambda_{0} =t^{2}-\left(  r-r'\right)^{2}\ ,
\end{equation}
\begin{equation}
\lambda_{\pi}  =t^{2}-\left(  r+r'\right)^{2} \ .
\end{equation}

In general, the field value given by the expression (\ref{Phiv}) can
be evaluated only numerically.  This evaluation, however, can be done
to a high precision very efficiently, for example, by using the
numerical integration package of the GNU Scientific Library
\cite{gsl}. 

\begin{figure}[htbp!]
 \centerline{
  \epsfxsize=8.5cm 
\epsfbox{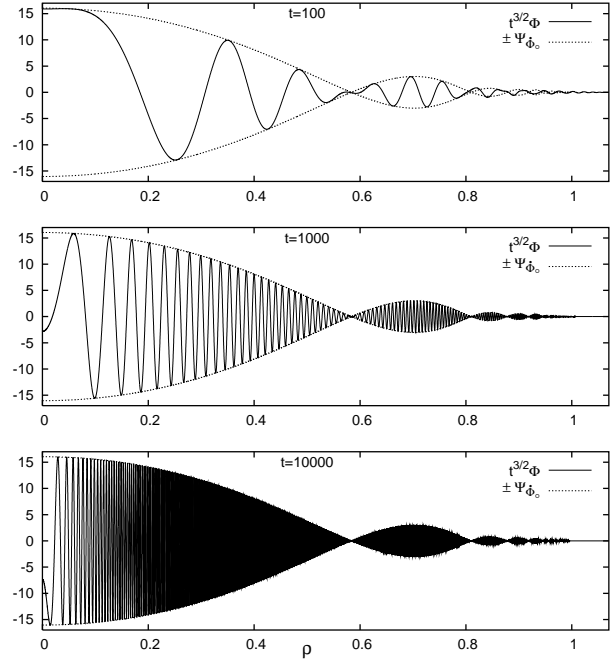}
 }
\caption{\footnotesize \label{fig1} The time evolution of a KG field
  with initial data $\Phi|_{\Sigma_0}=0$ and
  $(\partial_t\Phi)|_{\Sigma_0}=\dot{\Phi}_\circ$ is shown. In order to
  compensate the fast expansion and the decay in time, instead of
  the field  amplitude $\Phi$ the combination $t^{ 3/2}\Phi$ is
  plotted on the three successive time slices $t=100$, $t=1000$, and
  $t=10000$ against the variable $\rho=\frac rt$. 
  Notice that the value $\rho=1$ labels the point sitting at the light
  cone emanating from the center at $t=0$. The graphs
  $\pm\Psi_{\dot{\Phi}_\circ}$ -- for the definition of
  $\Psi_{\dot{\Phi}_\circ}$ see  (\ref{f1}) and (\ref{f11}) -- are
  also shown by the dashed curves. On the early time slice $t=100$ the
  final segment does not fit the contours
  $\pm\Psi_{\dot{\Phi}_\circ}$; however, at later times where
  conditions \ref{cond} and \ref{cond2} are satisfied in larger and
  larger portions of the time slices, the  exact oscillations get
  closer and closer to the contours. }
\end{figure}

\begin{figure}[htbp!]
 \centerline{
  \epsfxsize=8.5cm 
\epsfbox{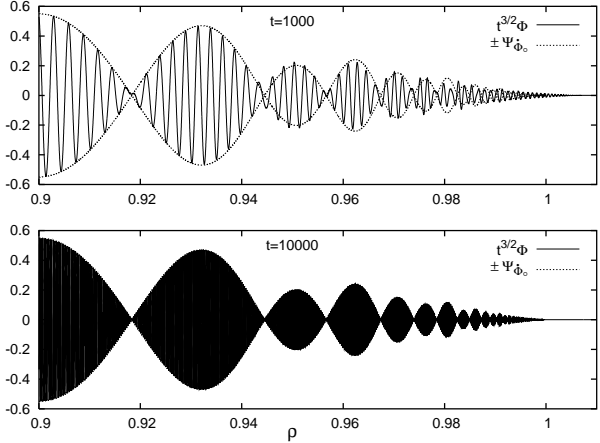}
 }
\caption{\footnotesize \label{fig1a} The final segments of the plots
  of Fig.\ \ref{fig1} for $\rho > 0.9$ on
  the time slices $t=1000$ and $t=10000$. It is remarkable
  to what extent the high frequency oscillations fit to the contours
  $\pm\Psi_{\dot{\Phi}_\circ}$ even in the region far away from the
  origin on the time slice $t=10000$.} 
\end{figure}

In the concrete examples described in detail in this paper the
initial data functions are chosen to possess the form 
\begin{equation}f(r) = \left\{ \begin{array} {rr} 
    c \exp\left[\frac{d}{(r-a)^2-b^2}\right], & {\rm if}\ r\in
    [a-b,a+b] ;\\  0 , & {\rm otherwise},  \end{array} \right.\label{ff}
\end{equation}
for $r\geq0$ with center at $r=a(\geq 0)$ and with width $2b> 0$,
which is a smooth function with compact support. In particular, in this
section the graphs presented refer to the evolution of an initial
data function $\dot{\Phi}_\circ=f(r)$ with $a=4$, $b=3$, $c=1$, and 
$d=10$. This initial data function  corresponds to ``hitting'' the 
vacuum configuration
between two concentric shells at $r=1$ and $r=7$ with a bell shape
distribution.  (The energy density and the energy current density
profiles associated with the same initial configuration are shown by 
Figs.\ \ref{fig6}, \ref{fig6a}, and \ref{fig7} below.)  The field  value
$\Phi$ associated with $m=1$ at 
the time level surfaces $t=100$, $t=1000$, and $t=10000$ are as
shown in Fig.\ \ref{fig1}. The final segment of the last two time
slices is plotted in Fig.\ \ref{fig1a}. In order to facilitate
comparison of the different time slices, to compensate the fast
expansion the field value is plotted as a function of the
self-similarity variable $\rho=\frac{r}{t}$. The striking similarity
of the structures at different time slices also holds for other
choices of initial data functions, although the exact shape and
appearance of the shell contours may differ significantly.

Next, we describe a method based on an approximation which explains
the appearance of these stable structures and provides a quantitative
account of the underlying phenomenon to a high precision. It is surprising by
itself that under the conditions introduced below the integrand  of
Eq.\ (\ref{Phiv}) can be very  precisely approximated by a function so that
the integral (\ref{Phiv}) takes a very simple form.

We denote by $\bar r'$ the radius of the smallest sphere
$\mathcal{B}(O,\bar r')$ centered at the origin so that
$\mathcal{B}(O,\bar r')$ contains the compact support of
the specified initial data on $\Sigma_0$. 

{\condition:\label{cond} We shall say that a point $p$, with coordinates
  $ (t,r)$, of the time slice $\Sigma_t$ is in the region of the 
  approximation if for a given $\epsilon<1$ the relation 
\begin{equation}
\bar r'< \epsilon(t-r)
\end{equation}
holds.}

Notice that on a fixed time slice the smaller the value of
$\epsilon$ the more precise 
the approximated field value, although the domain of the approximation
gets smaller. The numerical results indicate that even for the
relatively large $\epsilon=1/10$ there is surprisingly  good
agreement between 
the exact and approximated field values. For an arbitrarily small
fixed value of $\epsilon$ the domain where the approximation breaks
down is $t-\bar r'/\epsilon<r<t+ r'$. The size of this region,
independently of time, is
$\bar r'(1+1/\epsilon)$, while the domain of the
approximation $0\leq t-\bar r'/\epsilon$ grows linearly with $t$.  
Thereby one may think
of condition \ref{cond} as holding asymptotically almost everywhere
inside the causal future of the compact support of the initial data.
 
In the domain where condition \ref{cond} holds, both $\lambda_0$ and
$\lambda_\pi$ are positive, so then $H(\lambda_{0})=H(\lambda_{\pi})=1$.
In addition, we have that 
\begin{equation}
\frac{r' r}{t^2-r^2}=\frac{r' r}{(t-r)(t+r)}<\frac{r'}{(t-r)}< \epsilon
\end{equation}
and similarly
\begin{equation}
\frac{r'^2}{t^2-r^2}<\frac{r'}{(t-r)}\frac{r'}{t}< \epsilon.
\end{equation}
Then, by making use of these relations, along with the notation 
\begin{equation}
l_0=\sqrt{t^2-r^2},
\end{equation}
$\sqrt{\lambda_{0}}$ and $\sqrt{\lambda_{\pi}}$ can be
approximated as 
\begin{eqnarray}
\sqrt{\lambda_{0}}&\approx&
l_0\left(1+\frac{r' r}{l_0^2}-\frac{r'^2}{2l_0^2}\right), \label{l0ap} \\
\sqrt{\lambda_{\pi}}&\approx&
l_0\left(1-\frac{r' r}{l_0^2}-\frac{r'^2}{2l_0^2}\right).\label{lpiap}
\end{eqnarray}
Moreover, for $z\gg 1$ the Bessel function $J_0(z)$ can also be
very closely approximated as
\begin{equation}
J_{0}(z)\approx\sqrt{\frac{2}{\pi z}}\sin\left(z+\frac{\pi}{4}\right) .
\label{J0ap}
\end{equation}
By virtue of Eqs.\ (\ref{l0ap}) and (\ref{lpiap}), whenever condition
\ref{cond} holds, both the arguments of $J_{0}$ in Eq.\
(\ref{Phiv}), i.e., $m \sqrt{\lambda_{0}}$ and $m
\sqrt{\lambda_{\pi}}$, are much greater than $1$ provided that the 
following condition holds. 
{\condition:\label{cond2} The condition
\begin{equation}
m l_0\gg 1
\end{equation}
is satisfied.} 

This condition certainly holds in the region of approximation of
condition \ref{cond} if $t$
is chosen to be sufficiently large, i.e., for $t\gg \frac 1m$.  

For large values of $z$ the function $\sqrt{\frac{2}{\pi z}}
    \sin\left(z +\frac{\pi}{4}\right)$ has a high frequency 
    oscillation with a negligible change in the amplitude. Therefore
    we shall further simplify the approximation (\ref{J0ap}) of
    $J_{0}(z)$ applied 
    for the arguments $z=m\sqrt{\lambda_{0}}$ and $z=m
    \sqrt{\lambda_{\pi}}$ as  
\begin{equation}
J_{0}(z)\approx\sqrt{\frac{2}{m\pi l_0 }}\sin\left(z+\frac{\pi}{4}\right) .
\label{J0ap2}
\end{equation}
Hence, whenever conditions \ref{cond} and \ref{cond2} are satisfied,
the substitution of Eq.\ (\ref{J0ap2}) into Eq.\ (\ref{Phiv}) yields
\begin{eqnarray}
&& \hspace{-.7cm}\Phi(t,r)\approx
\frac{1}{2r}\sqrt{\frac{2}{\pi m\, l_0}}
\int\limits_{0}^{\infty}r' \dot{\Phi}_\circ(r')\nonumber\\ &&\hspace{.5cm} 
\cdot \left[
\sin\left(m\sqrt{\lambda_{0}}+\frac\pi4\right)
-\sin\left(m\sqrt{\lambda_{\pi}}+\frac\pi4\right)\right]dr'. 
\nonumber\\
\end{eqnarray}
Then by making use of Eqs.\ (\ref{l0ap}), (\ref{lpiap}) and keeping the
leading order term, a straightforward calculation results in  
\begin{eqnarray}
&& \hspace{-1.8cm}\Phi(t,r)\approx
\frac{1}{r}\sqrt{\frac{2}{\pi m\, l_0}}
\cos\left(ml_0+\frac\pi4\right) \nonumber\\
&&\hspace{.9cm}
\cdot \int\limits_{0}^{\infty}r' \dot{\Phi}_\circ(r')
\sin\left(\frac{m r}{l_0}r'\right)dr'.
\label{prtx}
\end{eqnarray}
Finally, by introducing $\rho=\frac{r}{t}$, the self-similarity
variable, Eq.\ (\ref{prtx}) can be recast into the form  
\begin{equation}
\Phi(t,r)\approx {t^{-\frac32}}\Psi_{\dot{\Phi}_\circ}(\rho)
\cos\left(m\sqrt{t^2-r^2}+\frac\pi4\right), \label{f1} 
\end{equation}
where
\begin{eqnarray}
\Psi_{\dot{\Phi}_\circ}(\rho)&=&\sqrt{\frac{2}{m \pi}}
\rho^{-1} (1-\rho^2)^{-\frac14}\nonumber\\ & &
\cdot \int\limits_{0}^{\infty}r' \dot{\Phi}_\circ(r')\sin\left(\frac{m\rho 
r'}{\sqrt{1-\rho^2}}\right)dr'. \nonumber\\ \label{f11} 
\end{eqnarray}
Notice that the integral term of Eq.\ (\ref{f11}) is, in fact, the real
part of the Fourier transform of the function $r\dot{\Phi}_\circ(r)$.
The cosine term of Eq.\ (\ref{f1}) gives a high frequency oscillation
that is completely independent of the specified initial data function,
i.e., of $\dot{\Phi}_\circ(r)$. The  amplitude of  this high
frequency oscillation is modulated by the rest of the
expression. There is an overall factor $t^{-3/2}$ scaling down the
modulation in time. However, the term $\Psi_{\dot{\Phi}_\circ}(\rho)$
depends  on $t$ and $r$ only in the combination $\rho={r}/{t}$, so this
term represents the self-similar outward expansion of the modulation.

For simple polynomial choices of the initial data function
$\dot{\Phi}_\circ$, the explicit form 
of $\Psi_{\dot{\Phi}_\circ}(\rho)$ can always be
determined.  In particular, for a centered step function, i.e., with
$\dot{\Phi}_\circ$ taking the value $c\in \mathbb{R}^+$ for $0\leq
r\leq r_b$ and being zero elsewhere, 
\begin{eqnarray}
\Psi_{\dot{\Phi}_\circ}(\rho)&=&c\sqrt{\frac{2}{m\pi}}\frac{(1-\rho^2)^\frac
34}{m^2\rho^3}\nonumber\\ & & \hspace{-0.5cm}
\cdot\left[\sin\left(\frac{m r_b \rho}{\sqrt{1-\rho^2}}\right)-
\frac{mr_b\rho}{\sqrt{1-\rho^2}}\cos\left(\frac{mr_b\rho}{\sqrt{1-\rho^2}}
\right)\right]. \nonumber\\\label{prt1}
\end{eqnarray}

\begin{figure}[htbp!]\label{figc}
 \centerline{
  \epsfxsize=8.5cm 
\epsfbox{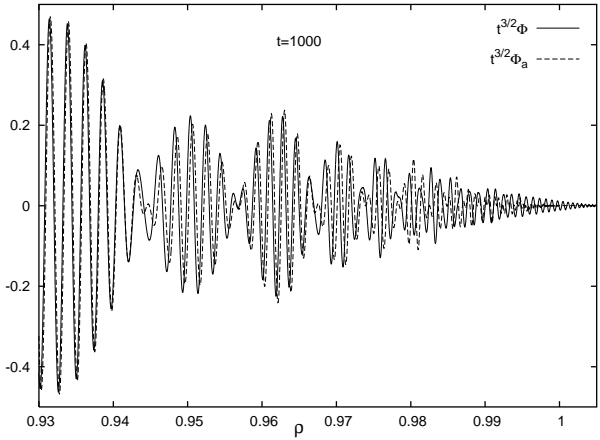}
 }
\caption{\footnotesize \label{fig4} The exact value field value
  $\Phi$ along with its value $\Phi_a$ as it is approximated by the
  relation (\ref{f1}) are shown for $\rho>0.93$ and
  for the time slice $t=1000$.}
\end{figure}

It is remarkable to what extent the high frequency oscillations fit
the contours $\pm\Psi_{\dot{\Phi}_\circ}$ (see Figs. \ref{fig1} and
\ref{fig1a}). We 
would like to emphasize, however, that not only this overall behavior
can be described properly by the approximation method outlined
above. Figure $3$ shows the exact field value $\Phi$, along with
its approximated value $\Phi_a$, where the values of $\Phi$ and
$\Phi_a$ are determined by making use of the relations (\ref{Phiv})
and (\ref{f1}), respectively. Since there is an almost exact
coincidence between the two relevant curves, only the final segment
with $\rho> 0.93$ is shown in Fig.\ \ref{fig4} for the time slice
$t=1000$.    

We would like to emphasize that for a fixed radius $r$ and for large
enough values of $t$ we have that  
$\Psi_{\dot{\Phi}_\circ}(\rho) \approx
\Psi_{\dot{\Phi}_\circ}(0)=\sqrt{\frac{2m}{\pi}}
\int\limits_{0}^{\infty}{r'}^2
\dot{\Phi}_\circ(r')dr' $ 
and $l_0 \approx t$, so that for a scalar monopole Eq.\ (\ref{f1})
restores the inverse power-law behavior of Eq.\ (\ref{af1})
relevant for $l=0$.

\begin{figure}[htbp!]
 \centerline{
  \epsfxsize=8.5cm 
\epsfbox{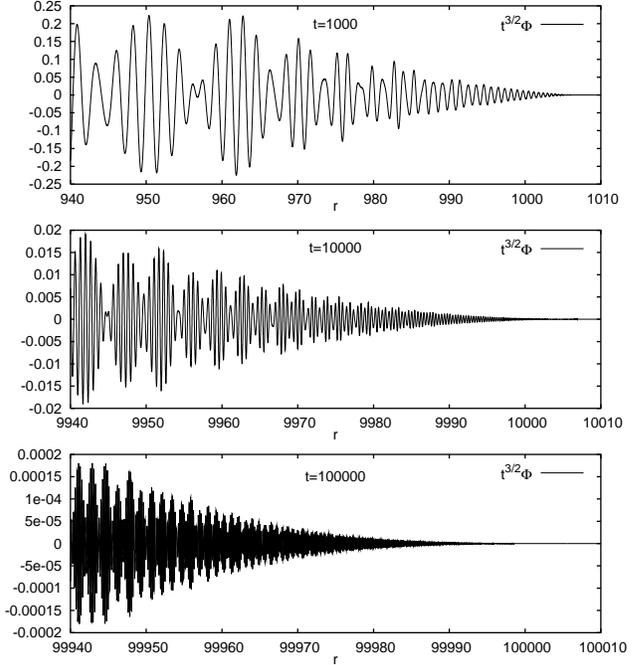}
 }
\caption{\footnotesize \label{figend} Equal length ($\Delta r=60$)
final segments of the true (nonapproximated) oscillations of 
$t^{-3/2}\Phi$ are shown at the time slices $t=1000$, $t=10000$, and
$t=100000$. The associated graphs demonstrate that the increase of the
maximal frequency of the oscillations does really occur, even though
the plots are made against the $r$ coordinate representing real
physical distances.}
\end{figure}

One might have the impression that the apparent increase of the
maximal frequency of the oscillations on the above figures is merely a
coordinate effect, i.e., it is a simple consequence of the use of the
self-similarity variable $\rho=r/t$, and one might suspect that it will
not occur if the plots are made against the radial coordinate
$r$. Nevertheless, the following graphs (see Fig.\ \ref{figend})
show that the increase of the maximal frequency
along the $t=const.$ slices is a true physical effect.

A simple explanation of this phenomenon can be given based on the use
of the approximation (\ref{f1}) as follows. Since the term responsible for
the high frequency oscillations is $\cos(ml_0(t,r)+\pi/4)$ the
frequencies $\omega_t$ and $\omega_r$ -- relevant for the $r=const.$
lines and for the $t=const.$ hypersurfaces -- can be read off from the 
approximation of the term $ml_0$ in the
neighborhood of a point with coordinates $(  t,  r)$ as
\begin{eqnarray}
\hskip -.4cm ml_0(  t+\Delta t,  r+\Delta r) \approx ml_0( 
t,  r)+\omega_{  t}\Delta 
t-\omega_{  r}\Delta r ,
\end {eqnarray}
where the frequencies are
\begin{eqnarray}
\omega_{  t}&=&m\partial_t l_0=m\frac{  t} {\sqrt{{ 
      t}^2-{  r}^2}}, \label{omt}\\ 
\omega_{  r}&=&-m\partial_r l_0=m\frac{  r}{\sqrt{{ 
      t}^2-{  r}^2}},\label{omr}  
\end {eqnarray}
respectively. It follows from these expressions that as $r$ tends
to the value of $t$ both $\omega_t$ and $\omega_r$ grow
without any upper  bound, although their ratio tends to $1$. This, in
particular, implies that the self-similar part of the modulation
expands (at least at the ``outer edge'') with the velocity of light.
The frequency of the oscillation at a fixed distance $\Delta r=t-r$
from the ``outer edge'' ($t=r$) grows as $\sqrt{t}$, as it can be seen
from the approximation
\begin{equation}
\omega_r\approx\omega_t=\frac{t}{\sqrt{\Delta r}\sqrt{t+r}}
\approx\frac{1}{\sqrt{2\Delta r}}\sqrt{t} 
\end{equation}
relevant for $\Delta r\ll t$.
This, in turn, implies -- in accordance with Fig.\ \ref{figend} --
that more and more oscillations will be associated with the final
segments of length $\Delta r$ as the time increases.

It is also informative to consider what can be seen by a static
observer, moving along an $r=const.$ world-line with $r\gg \bar r'$.
First, a very high frequency oscillation arrives at $t=r-\bar r'$ with
amplitude growing gradually from zero.  In the domain where the
approximation (\ref{f1}) is valid, the amplitude of this  oscillation
becomes modulated by ${t^{-\frac32}}\Psi_{\dot{\Phi}_\circ}(\rho)$,
while $\rho=r/t$ decreases from a value close to $1$ to $0$.
By virtue of Eq.\ (\ref{omt})
the frequency of the high frequency oscillation gradually
decreases, settling down finally to the value $\omega_t=m$ in
accordance with Eq.\ (\ref{af1}).

\section{Evolution of a temporarily static configuration}\label{nvac}
\setcounter{equation}{0}

\begin{figure}[htbp!]
 \centerline{
  \epsfxsize=8.5cm 
\epsfbox{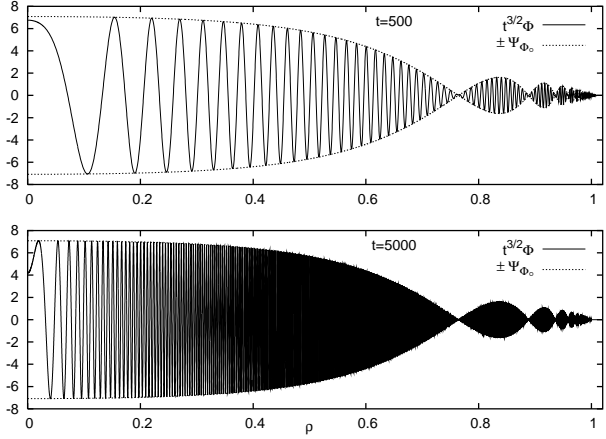}
 }
\caption{\footnotesize \label{fig5} The time evolution of a KG field
  with initial data function $\Phi|_{\Sigma_0}=\Phi_\circ$ and
  $(\partial_t\Phi)|_{\Sigma_0}=0$ is plotted at 
  the two successive time slices $t=500$ and $t=5000$ against the
  variable $\rho=\frac rt$. The graphs $\pm\Psi_{\Phi_\circ}$ -- for
  the definition of $\Psi_{\Phi_\circ}$ see Eqs.\ 
  (\ref{f2}) and (\ref{f22}) -- are also indicated by the dashed 
  curves.}
\end{figure}

\begin{figure}[htbp!]
 \centerline{
  \epsfxsize=8.5cm 
\epsfbox{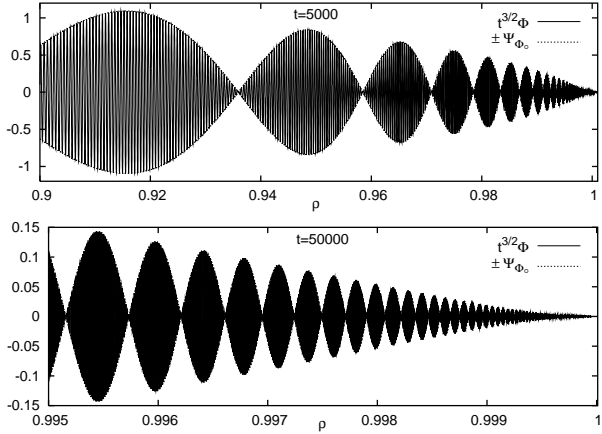}
 }
\caption{\footnotesize \label{fig5a} The final segment of the second plot
  of Fig.\ \ref{fig5} restricted to $\rho>0.9$ together with an even smaller
  portion relevant for a later time slice $t=50000$ with $\rho>0.995$
  are shown.}  
\end{figure}

Start now with a static initial data function, i.e., assume that
$(\partial_t\Phi)|_{\Sigma_0}=0$ and $\Phi|_{\Sigma_0}= \Phi_\circ(r)$ 
where
$\Phi_\circ:[0,\infty)\rightarrow \mathbb{R}$ is a sufficiently regular
function of $r$ with compact support. Now to get the associated field 
values we have to evaluate only the second term of Eq.\ (\ref{grth}). 
In particular, at a point of the axis of rotation with coordinates
$x^{\alpha}=(t,r,0,0)$, with $t>0$, we obtain from Eq.\ (\ref{grth})
\begin{eqnarray}
 \hspace{-.3cm}\Phi(t,r) &=& 
-\int\limits_{0}^{\infty}dr'\int\limits_{0}^{\pi}%
d\vartheta'\int\limits_{0}^{2\pi} d\varphi'\nonumber\\ && \cdot
\Phi_\circ(r') \left[\frac{\partial}{\partial {t'}}  
G(t,r;t',r',\varphi',\vartheta')\right]_{t'=0}
\hspace{-.2cm}{r'}^{2}\sin\vartheta'  
\nonumber\\ 
\hspace{-.3cm}&=&-\frac{t}{r}\int\limits_{0}^{\infty}dr'
r'\Phi_\circ(r')\int\limits_{0}^{\pi} d\vartheta' \nonumber\\ && \cdot
\frac{\partial}  
{\partial\vartheta'}\left[  \delta(\lambda) -\frac{m}{2\sqrt{\lambda}}
H(\lambda)J_{1}(m\sqrt{\lambda})\right],
\label{p2}
\end{eqnarray}
where the relation 
\begin{equation}
\frac{\partial}{\partial {t'}}=\frac{\partial \lambda}{\partial {t'}}
\left(\frac{\partial \lambda}{\partial {\vartheta'}}\right)^{-1}
\frac{\partial }{\partial {\vartheta'}}= \frac{t-t'}{r' r \sin \vartheta'}
\frac{\partial }{\partial {\vartheta'}}, \label{gf2t}
\end{equation}
along with Eq.\ (\ref{gf1}), has also been used. By evaluating the
$\vartheta'$ integral of Eq.\ (\ref{p2}) we get
\begin{eqnarray}
& &\hspace{-.6cm}\Phi(t,r) =
\frac{t}{r}\int\limits_{0}^{\infty}dr'\ r' {\Phi}_\circ(r')
\bigg[\delta(\lambda_0) -\delta(\lambda_\pi)
\bigg. \nonumber\\
& &  
\hspace{-.2cm} \bigg.-\frac{m}{2}
\left( \frac{H(\lambda_0)}{\sqrt{\lambda_0}} 
J_{1}(m\sqrt{\lambda_0})-\frac{H(\lambda_\pi)}{\sqrt{\lambda_\pi}} 
J_{1}(m\sqrt{\lambda_\pi})\right)\bigg]. 
\nonumber\\
\label{Phivt}
\end{eqnarray}
For determination of the parts of the above integral containing
the Dirac delta terms we used the relations  
\begin{eqnarray}
\int\limits_{0}^{\infty}\delta(\lambda_0)\psi(r')dr'
&=&\int\limits_{0}^{r}
\delta(\lambda_0)\psi(r')dr' +
\int\limits_{r}^{\infty}\delta(\lambda_0) \psi(r')dr' \nonumber\\
&=&\int\limits_{t^2-r^2}^{t^2}
\delta(\lambda_0)\psi(r')(\partial_{r'}\lambda_0)^{-1} d\lambda_0
\nonumber\\  
&+&\int\limits_{t^2}^{-\infty}\delta(\lambda_0)
\psi(r')(\partial_{r'}\lambda_0)^{-1} d\lambda_0
\nonumber\\
&&\hspace{-2.4cm}=\frac12 H(r-t)\left(\frac{\psi(r')}{r-r'}
\right)_{r'=r-t}
-\frac12\left(\frac{\psi(r')}{r-r'} \right)_{r'=r+t}  \nonumber\\
&&\hspace{-2.4cm}=\frac{1}{2t}\big[\psi(r+t)+H(r-t)\psi(r-t) \big],
\end{eqnarray}
and
\begin{eqnarray}
\hspace{-.3cm}\int\limits_{0}^{\infty}\delta(\lambda_\pi)\psi(r')
 dr'&=&\frac12 
 H(t-r) \left(\frac{\psi(r')}{r'+r} \right)_{r'=t-r}\nonumber\\
\hspace{1.8cm}&=&\frac{1}{2t} H(t-r) \psi(t-r), 
\end{eqnarray}
where $\psi:[0,\infty)\rightarrow \mathbb{R}$ is considered to be a
sufficiently regular but otherwise arbitrary function. Thereby, we get
from Eq.\ (\ref{Phivt}) 
\begin{eqnarray}
&& \hspace{-.9cm}\Phi(t,r) =-\frac{mt}{2r}\int\limits_{0}^{\infty}r'
  \Phi_\circ(r') \nonumber\\ &&  \hspace{-.1cm}\cdot
\left[\frac{H(\lambda_0)}{\sqrt{\lambda_0}} 
J_{1}(m\sqrt{\lambda_0}) 
-\frac{H(\lambda_\pi)}{\sqrt{\lambda_\pi}} 
J_{1}(m\sqrt{\lambda_\pi}) \right]dr'  \nonumber\\
& &\hspace{.6cm}+\frac{1}{2r}\big[(r+t)\Phi_\circ(r+t)
  +(r-t)\Phi_\circ(|r-t|) \big].\nonumber\\ 
\label{Phivt2}
\end{eqnarray}
This integral has been evaluated numerically and the relevant plots on
the time slices $t=500$ and $t=5000$ are shown in Figs.\ \ref{fig5} and
\ref{fig5a}. Here the initial data function was chosen to possess the 
analytic form of (\ref{ff}), i.e., $\Phi_\circ=f(r)$, with parameters
$a=0$, $b=5$, $c=1$ and $d=15$ corresponding to a bell shaped initial
configuration centered at the origin with radius $r=5$.   

In the domain where conditions \ref{cond} and \ref{cond2} are valid,
in particular, whenever $t$ is sufficiently large, the nonintegral
terms of Eq.\ (\ref{Phivt2}) are zero, since both $r+t$ and $|r-t|$ are
larger than the radius $\bar r'$ of the initial data function. 
Furthermore, the integral term in Eq.\ (\ref{Phivt2}) can be 
approximated by exactly the same procedure as that used in case 
of Eq.\ (\ref{Phiv}). To see this, note
first that for $z\gg 1$ the Bessel function $J_1(z)$ can be very
closely approximated as
\begin{equation}
J_{1}(z)\approx-\sqrt{\frac{2}{\pi z}}\cos\left(z+\frac{\pi}{4}\right).
\label{J1ap}
\end{equation}
Then, in particular, whenever conditions \ref{cond} and \ref{cond2} are
satisfied by making use of Eqs.\ (\ref{l0ap}) and (\ref{lpiap}), 
along with the
self-similarity variable $\rho=\frac{r}{t}$, the integral term, and
thereby $\Phi(t,r)$ itself, can be approximated as
\begin{equation}
\Phi(t,r)\approx
{t^{-\frac32}}\Psi_{\Phi_\circ}(\rho)\sin\left(m\sqrt{t^2-r^2}
+\frac\pi4\right), \label{f2} 
\end{equation}
where
\begin{eqnarray}
&& \hspace{-.8cm}\Psi_{\Phi_\circ}(\rho)=-\sqrt{\frac{2m}{\pi}}
\rho^{-1} (1-\rho^2)^{-\frac34}\nonumber\\ &&\hspace{1.8cm}
\cdot\int\limits_{0}^{\infty}r' \Phi_\circ(r')\sin\left(\frac{m\rho 
r'}{\sqrt{1-\rho^2}}\right)dr'. \nonumber\\ \label{f22} 
\end{eqnarray}
Notice that Eq.\ (\ref{f2}) has the same type of structure as Eq.\
(\ref{f1}). The sine term of Eq.\ (\ref{f2}) again represents a high
frequency oscillation that is completely independent of the
specified initial data. The amplitude of 
this high frequency oscillation is modulated by the overall factor
$t^{-3/2}$, scaling down the 
modulation in time, along with $\Psi_{\Phi_\circ}$. Again the term
$\Psi_{\Phi_\circ}$ depends on 
$t$ and $r$ only via the combination $\rho={r}/{t}$ and as in the 
case of Eq.\ (\ref{f1}) this term gives a self-similar outward 
expansion of the modulation.

\section{The appearance of the shells in terms of the energy
  density}\label{en} 
\setcounter{equation}{0}

\begin{figure}[bp!]
 \centerline{
  \epsfxsize=8.5cm 
\epsfbox{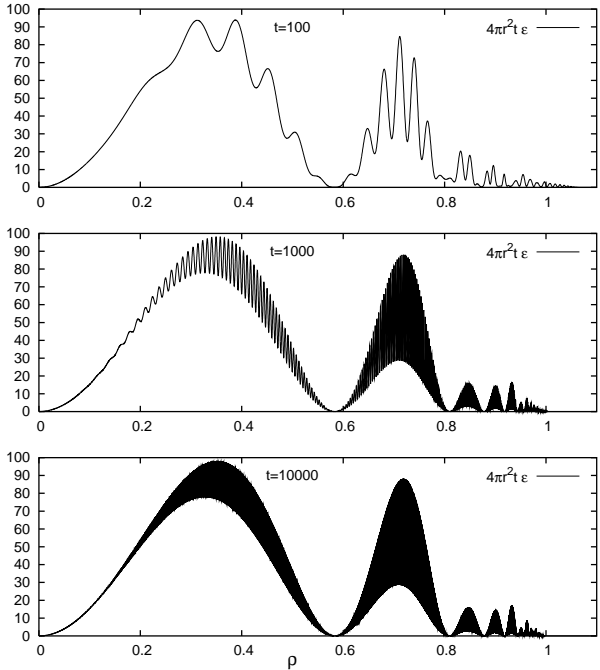}
 }
\caption{\footnotesize \label{fig6} The energy density profile
  relevant for the excitation of the initially vacuum 
  configuration of Sec.\ \ref{vac}, is shown
  for the time slices $t=100$, $t=1000$ and $t=10000$.}
\end{figure}

\begin{figure}[htbp!]
 \centerline{
  \epsfxsize=8.5cm 
\epsfbox{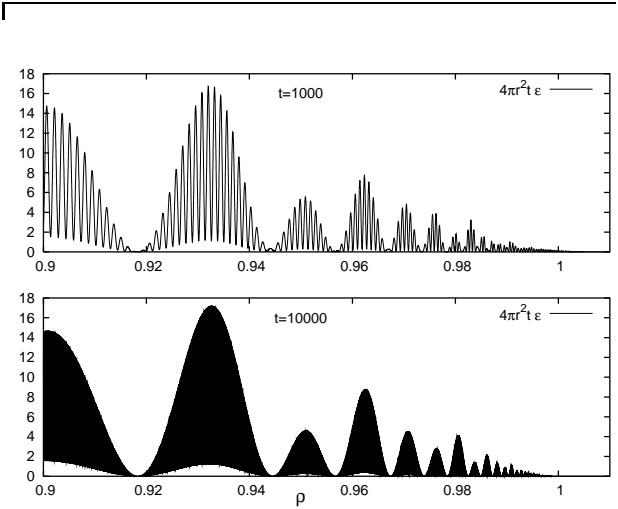}
 }
\caption{\footnotesize \label{fig6a} The final segments of the energy
  density profile of Fig.\ \ref{fig6} is shown for $\rho>0.9$ and 
  for the time slices $t=1000$ and $t=10000$.}
\end{figure}

As is obvious from the relations (\ref{ed}) and (\ref{ec}), to be able
to evaluate the energy density and the energy current density
expressions in addition to Eqs.\ (\ref{Phiv}) and (\ref{Phivt2}) we 
also need to
determine the derivatives of the field variable $\Phi$ with respect to
$t$ and $r$, respectively. These derivatives, by straightforward
calculation, although a bit more involved than the previous ones, for a
generic initial data $\Phi|_{\Sigma_0}=\Phi_\circ(r)$ and
$(\partial_t\Phi)|_{\Sigma_0}=\dot{\Phi}_\circ(r)$ can be  shown to
take the form
\begin{widetext}
\begin{eqnarray}
\partial_t \Phi(t,r) &=& \frac{1}{t} \Phi(t,r) - \frac{mt}{2r}
\int\limits_{0}^{\infty} r'\dot{\Phi}_\circ(r')\left[
\frac{H(\lambda_{0})}{\sqrt{\lambda_{0}}}J_{1}(m\sqrt{\lambda_{0}})-
\frac{H(\lambda_{\pi})}{\sqrt{\lambda_{\pi}}}J_{1}(m\sqrt{\lambda_{\pi}})
\right]dr' \nonumber\\ & & 
\hspace{1.1cm}
+\frac{m^2t^2}{2r}\int\limits_{0}^{\infty} r'\Phi_\circ(r')\left[
\frac{H(\lambda_{0})}{\lambda_{0}}J_{2}(m\sqrt{\lambda_{0}})
-\frac{H(\lambda_{\pi})}{\lambda_{\pi}}J_{2}(m\sqrt{\lambda_{\pi}})
\right]dr' \nonumber\\ & &
+\frac{1}{2r}\big[(r+t)\dot{\Phi}_\circ(r+t)+(r-t)\dot{\Phi}_\circ(|r-t|)+
  (r+t)\Phi_\circ'(r+t)-|r-t|\Phi_\circ'(|r-t|) 
\big]\nonumber\\ & & 
-\frac{1}{2t}\big[\Phi_\circ(r+t)-\Phi_\circ(|r-t|)\big] - \frac{m^2t}{4r}
\big[(r+t)\Phi_\circ(r+t)+(r-t)\Phi_\circ(|r-t|) \big]
\label{dtPhiv}  
\end{eqnarray}
and
\begin{eqnarray}
\partial_r \Phi(t,r) &=&
+\frac{m}{2r}\int\limits_{0}^{\infty} r'\dot{\Phi}_\circ(r') \left[
\frac{H(\lambda_{0})(r-r')}{\sqrt{\lambda_{0}}}J_{1}(m\sqrt{\lambda_{0}})
-\frac{H(\lambda_{\pi})(r+r')}{\sqrt{\lambda_{\pi}}}
J_{1}(m\sqrt{\lambda_{\pi}}) \right]dr'\nonumber\\ & &
-\frac{m^2t}{2r}\int\limits_{0}^{\infty} r'\Phi_\circ(r') \left[
\frac{H(\lambda_{0})(r-r')}{\lambda_{0}}J_{2}(m\sqrt{\lambda_{0}})
-\frac{H(\lambda_{\pi})(r+r')}{\lambda_{\pi}}
J_{2}(m\sqrt{\lambda_{\pi}})   
\right]dr'\nonumber\\& & 
- \frac{m^2t}{4r} \big[(r+t)\Phi_\circ(r+t)-(r-t)\Phi_\circ(|r-t|) \big]
+\frac{1}{2r}\big[(r+t)\dot{\Phi}_\circ(r+t)-(r-t)\dot{\Phi}_\circ(|r-t|)\big.
\nonumber\\& & 
\big.
+\Phi_\circ(r+t)+(r+t)\Phi_\circ'(r+t)+
\Phi_\circ(|r-t|)+|r-t|\Phi_\circ'(|r-t|)\big]-\frac{1}{r} \Phi(t,r),\nonumber\\
\label{drPhiv}  
\end{eqnarray}
\end{widetext}
where $\Phi_\circ'$ stands for the $r$ derivative of the function 
$\Phi_\circ(r)$. Then by making use of Eqs.\ (\ref{ed}) 
and (\ref{ec}), along with Eqs.\ (\ref{Phiv}), (\ref{Phivt2}),
(\ref{dtPhiv}), and (\ref{drPhiv}), we can evaluate the energy density
and energy current density expressions. For the choice of initial data
function used in Sec.\ \ref{vac} to produce Figs.\ \ref{fig1} and 
\ref{fig1a}, the various plots  
of the energy density on the time level surfaces $t=100$, $t=1000$, and
$t=10000$ are shown in Figs.\ \ref{fig6} and \ref{fig6a}. 

\begin{figure}[htbp!]
 \centerline{
  \epsfxsize=8.5cm 
\epsfbox{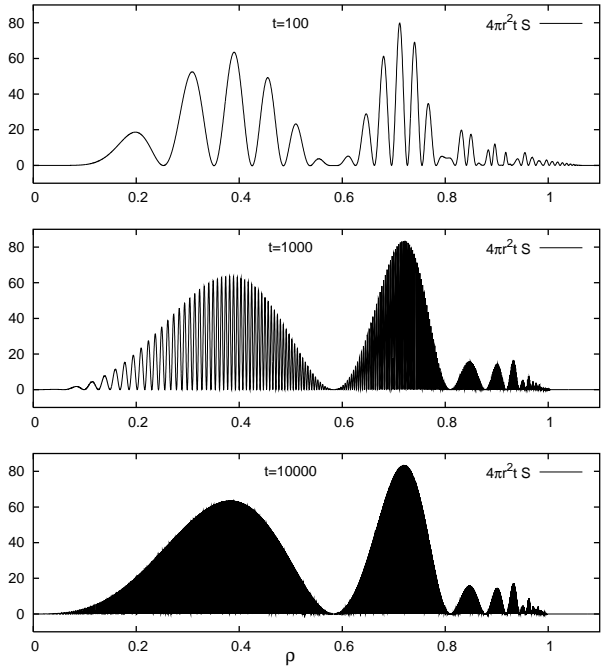}
 }
\caption{\footnotesize \label{fig7} The energy current density
  profile relevant for the excitation of the initially vacuum
  configuration of Sec.\ \ref{vac} is shown
  for the time slices $t=100$, $t=1000$, and $t=10000$.}
\end{figure}

The corresponding energy current density profile on the time level
surfaces $t=100$, $t=1000$ and 
$t=10000$ is shown in Fig.\ \ref{fig7}. 

\begin{figure}[htbp!]
 \centerline{
  \epsfxsize=8.5cm 
\epsfbox{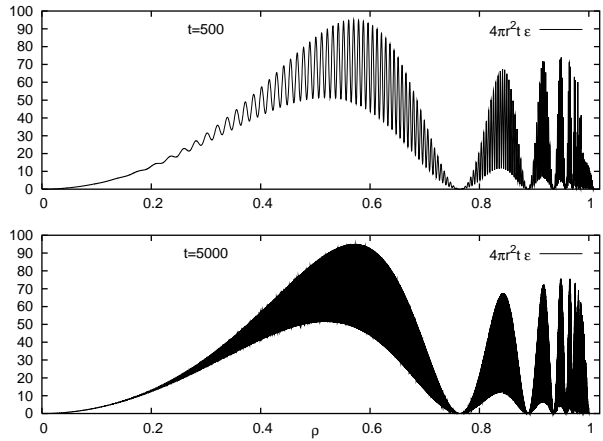}
 }
\caption{\footnotesize \label{fig8} The energy density profile
  relevant for the excitation of the initially static configuration of
  Sec.\ \ref{nvac}, is shown for the time slices $t=500$ and $t=5000$.}
\end{figure}

\begin{figure}[htbp!]
 \centerline{
  \epsfxsize=8.5cm 
\epsfbox{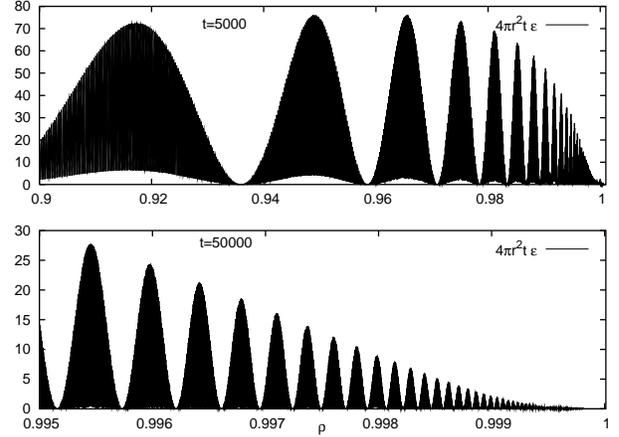}
 }
\caption{\footnotesize \label{fig8a} Two different final segments of 
  the energy density profile of Fig.\ \ref{fig8} are shown 
  for the time slice $t=5000$ with $\rho>0.9$ and for the time slice
  $t=50000$ with $\rho>0.995$.}
\end{figure}

Similarly, for the choice of initial data function used in 
Sec.\ \ref{nvac} to
produce Figs.\ \ref{fig5} and \ref{fig5a}, the various plots of the
energy density on the time level surfaces $t=500$, $t=5000$, and 
$t=50000$ are shown in Figs.\ \ref{fig8} and \ref{fig8a}. 

In order to represent more clearly the ratio of the total energy
carried by individual shells in the spherically symmetric
configuration investigated, instead of the energy density 
$\varepsilon$ and the
energy current density $S$, we plotted the quantities $4\pi
r^2\varepsilon$ and $4\pi r^2S$ representing the energy and energy
current in thin spherical shells. Since we are using $\rho=r/t$ as a
radial coordinate on our plots we also have to multiply these values
by $\frac{dr}{d\rho}=t$ 
\begin{equation}
E=\int_0^{t+\bar r'}\hspace{-.15cm}  4\pi r^2\varepsilon dr= 
\int_0^{1+\frac{\bar r'}{t}}\hspace{-.15cm}  4\pi r^2 t\varepsilon
d\rho ,
\end{equation}
in order to be able to estimate the energy as the area under the
plotted curves. 

Notice that the formation of self-similarly expanding shells is 
manifested strikingly by Figs.\ \ref{fig6} - \ref{fig8a} in spite 
of the fact that
no approximation has been used anywhere in the associated
calculations. 

\section{Concluding remarks}\label{fine}
\setcounter{equation}{0}

In this paper the time evolution of initially concentrated spherically
symmetric massive KG fields has been investigated. By means of concrete
examples and by a  suitable approximation method the following
characteristic properties have been found. There is an overall high
frequency oscillation of the field value which is modulated by two
factors. First, there is a time decaying factor of the form $t^{-3/2}$
consistent with the total energy conservation.  Second, the field
amplitude is also modulated by a self-similar contour, i.e., apart from
the scaling down $t^{-3/2}$ the oscillations expand essentially
self-similarly. The associated self-similar contours were found to
possess knots, whereby the oscillations are separated into individual
shells. Since the overall exponent of the scaling down factor is
independent of the radial coordinate, the fraction of the energy
carried by an individual expanding shell is also preserved during the
entire evolution. The exact form of the shells, which is approximated
to higher and higher precision in larger and larger domains of the
associated domain of dependence, is determined essentially by the
Fourier transform of the initial data.

It is of obvious interest to know whether the results presented in
this paper have any relevance for initial data with noncompact
support. To be able to answer this question it is necessary to check
whether the conditions applied throughout our analysis,
i.e., conditions \ref{cond} and \ref{cond2}, can be satisfied at least
in a generalized sense. We have investigated this issue briefly and we
found the following. Suppose that we have a non-compact support 
initial data function, which, however, is focused sufficiently on the 
central region, i.e., it decays rapidly. Then we can associate a 
finite characteristic size with the support of such an initial data
function. We expect that condition \ref{cond} can be
replaced by the requirement that this characteristic size should be 
small compared to $\epsilon(t-r)$. If we have an initial data function 
of this type and,
in addition, condition \ref{cond2} holds, basically the same type of
approximations and 
plots can be produced for the field values and for the energy
densities as was possible in the case of initial data with compact
support. It might be
interesting that for initial data functions 
$\Phi|_{\Sigma_0}=\Phi_\circ(r)$ and
$(\partial_t\Phi)|_{\Sigma_0}=0$, where  $\Phi_\circ(r)$ was chosen to
possess the form $\Phi_\circ(r)=\exp(-r^2/d)$ (with
$d\in \mathbb{R}^+$), there was no knot on the the associated contours
$\pm\Psi_{\Phi_\circ}$, while 
for the choice $\Phi_\circ(r)={d}/{(d+r^8)}$ (with $d\in \mathbb{R}^+$)
the very same type of figures, with shells and oscillations, were
obtained as in the case of initial data with compact support. 
Investigation of more general initial data specifications with
noncompact support, as well as
a careful adaptation of the approximation procedure applied throughout
this paper, deserves further attention. 

It would also be important to know to what extent the results
presented in this paper may have
relevance in more complicated physical situations, for instance, in
the case of nonlinear self-interacting fields, or whenever instead of 
the Minkowski spacetime the evolution is investigated on a general
asymptotically flat background spacetime. Note, however, that in the
asymptotic region the field equation relevant for a self-interacting
possibly nonlinear scalar field in many cases is expected to reduce to
the  equation of a massive KG field;
moreover, in that region the geometry also tends to the form of the
Minkowski metric. Thereby it seems to be plausible to assume that the
formation of shell structures, as well as the time  decay rate and
the self-similar behavior of the modulation, will probably  occur in
any asymptotically flat spacetime (regardless of whether the geometry 
is dynamical or not) and for any massive field that can be approximated
by a massive KG field in the asymptotic region. 

\section*{Acknowledgments}

The authors wish to thank P\'{e}ter Forg\'{a}cs and J\"{o}rg
Frauendiener for stimulating discussions. This research was supported
in parts by OTKA grant T034337 and NATO grant PST.CLG.978726. I.R. 
would like to thank the Bolyai Foundation for financial support.

\vfill\eject
\end{document}